# Protecting DeFi Platforms against Non-Price Flash Loan Attacks

Abdulrahman Alhaidari
Informatics and Networked Systems
University of Pittsburgh
Pittsburgh, PA, USA
aba70@pitt.edu

Balaji Palanisamy
Informatics and Networked Systems
University of Pittsburgh
Pittsburgh, PA, USA
bpalan@pitt.edu

Prashant Krishnamurthy
Informatics and Networked Systems
University of Pittsburgh
Pittsburgh, PA, USA
prashk@pitt.edu

## ABSTRACT

Smart contracts in Decentralized Finance (DeFi) platforms are attractive targets for attacks as their vulnerabilities can lead to massive amounts of financial losses. Flash loan attacks, in particular, pose a major threat to DeFi protocols that hold a Total Value Locked (TVL) exceeding $106 billion. These attacks use the atomicity property of blockchains to drain funds from smart contracts in a single transaction. While existing research primarily focuses on price manipulation attacks, such as oracle manipulation, mitigating non-price flash loan attacks that often exploit smart contracts' zero-day vulnerabilities remains largely unaddressed. These attacks are challenging to detect because of their unique patterns, time sensitivity, and complexity. In this paper, we present FlashGuard, a runtime detection and mitigation method for non-price flash loan attacks. Our approach targets smart contract function signatures to identify attacks in real-time and counterattack by disrupting the attack transaction atomicity by leveraging the short window when transactions are visible in the mempool but not yet confirmed. When FlashGuard detects an attack, it dispatches a stealthy dusting counterattack transaction to miners to change the victim contract's state which disrupts the attack's atomicity and forces the attack transaction to revert. We evaluate our approach using 20 historical attacks and several unseen attacks. FlashGuard achieves an average real-time detection latency of 150.31ms, a detection accuracy of over 99.93%, and an average disruption time of 410.92ms. FlashGuard could have potentially rescued over $405.71 million in losses if it were deployed prior to these attack instances. FlashGuard demonstrates significant potential as a DeFi security solution to mitigate and handle rising threats of non-price flash loan attacks.

## 1 INTRODUCTION

Smart contracts in blockchains manage decentralized finance (DeFi) protocols and enable users' financial transactions [4]. They autonomously enforce agreements between users and facilitate services, such as lending, asset management, and liquidity pool aggregation [39]. Even though smart contracts have become common in many financial interactions in DeFi, they are still prone to numerous vulnerabilities [6, 15] including vulnerabilities that have led to massive financial losses. The Total Value Locked (TVL) in DeFi is estimated to be $106 billion in 2024 [21], and is projected to expand further. However, as DeFi continues to grow, attacks have been rapidly increasing to exploit DeFi protocol vulnerabilities [60]. The total assets lost from these attacks amount to $79.8 billion of which only $6.7 billion has been recovered [20].

In this paper, we focus on flash loans, a temporary loan that is borrowed and repaid within one transaction block on the DeFi platform. Most flash loan providers do not require paying premiums because of the existence of massive liquidity vaults [57] which facilitate borrowing large amounts of assets, and guarantee that the borrowed funds are returned as enforced by the blockchain atomicity property. Atomicity in DeFi platforms ensures that all operations in a transaction either succeed entirely or are reverted without partial execution. If the fund is not returned at the block confirmation time, all internal transactions (actions) that depend on the flash loan are reverted. Blockchain atomicity has also been used to exploit DeFi vulnerabilities [41]. Flash loans were first used to manipulate the token price of DeFi platforms [41]. From there, new variations of large asset manipulations have emerged. In total, flash loan attacks have caused a total loss of more than $968 million [20].

Non-price flash loan attacks refer to a new variation of flash loan attacks where attackers bypass logical preconditions (i.e., state validation checks) to manipulate contract states without altering token prices [8]. This is contrast to traditional flash loan attacks that are usually associated with manipulating prices, either to inflate or deflate them [55]. Non-price flash loan attacks are not related directly to token prices but they share some properties with traditional flash loan attacks, such as atomic execution within a single transaction. Mitigating non-price flash loan attacks is inherently challenging because they mostly exploit zero-day vulnerabilities and usually have unique attack patterns, which make detection difficult. In addition, once assets are compromised, they cannot be recovered after a transaction is confirmed. For instance, in April 2024, due to an exploit of an unknown smart contract vulnerability in Hedgey Finance, the platform lost $44 million [20] that could not be recovered because the vulnerability was previously unknown to the platform.

In general, non-runtime defense solutions may fail to mitigate such non-price flash loan attacks, and we need effective countermeasures that can stop the attack by intervening in real-time. Previous research has mainly focused on flash loan attacks that manipulate prices, such as post-attack analysis and machine learning models for detection. These methods help to understand the attack vector, but they are not tailored for non-price flash loan attacks. Some real-time methods, such as STING [59] target general smart contract vulnerabilities, but they are insufficient for these attacks. Other studies, such as Flashot [13], propose tools to illustrate asset flows, but they do not provide detection or mitigation methods. AI-based approaches by analyzing transaction data achieve high accuracy, however, they do not support real-time intervention that is highly crucial for time-sensitive attacks involving flash loans [34]. FrontDef [22] is a mitigation system designed to synthesize counterattacks for suspicious pending transactions. However, FrontDef relies on high gas fees and has risks of transaction deadlock in public mempools that make it impractical for many scenarios. The above studies have key limitations in two ways: they either focus on general attacks or they do not support real-time mitigation that is crucial for stopping non-price flash loan attacks.

In this paper, we introduce a novel framework called FlashGuard for supporting runtime detection and defense that addresses existing research gaps in mitigating non-price flash loan attacks. FlashGuard disrupts attacks using a stealthy counterattack transaction by leveraging private relays [25] to bypass the public mempool and avoid detection or front-running. It interacts with the blockchain mempool to support real-time detection and effectively disrupts the flash loan attack by interacting with the DeFi victim's smart contract. Since FlashGuard continuously scans pending transactions to identify attack patterns, it dispatches a counterattack transaction once a possible attack is detected. The counterattack disrupts the atomicity of the non-price flash loan attack by altering the DeFi protocol's current state after the attacker calls the DeFi contract before it is confirmed. This prevents the attack transaction from being finalized and stops the attack from completing and mitigates its threat without requiring the redeployment of the current DeFi smart contract. FlashGuard is highly effective against non-price flash loan attacks and is generalizable to any EVM-based blockchain.

We evaluate FlashGuard on 20 real-world historical attacks that exploited DeFi protocol vulnerabilities from different blockchains such as Ethereum (ETH), Base (Base), Arbitrum (ARBI), Polygon (POLY), and Avalanche (AVAX). In addition, we examine six potential new attack scenarios to assess our method's adaptability to new vulnerabilities. Our experiments indicate that FlashGuard could have potentially rescued losses of over $405.71 million in DeFi platforms if it were deployed prior to these attacks. FlashGuard effectively provides an additional layer of protection for actively disrupting on-chain and real-time attacks to reduce financial losses and retain users' trust in DeFi platforms.

**Our Contributions.** Our paper makes the following contributions to safeguarding DeFi protocols from non-price flash loan attacks:

- **Real-Time Defense:** We propose FlashGuard, a real-time solution for detecting and disrupting non-price flash loan attacks. FlashGuard is designed to safeguard DeFi protocols without changing the current deployments.
- **Heuristic-Based Detection:** We develop a novel detection algorithm that monitors the mempool for non-price flash loan attack signatures and enables to track function signatures for each internal transaction in atomic transactions. It also distinguishes malicious transactions before they are included in the blockchain.
- **Atomicity Attack Disruption:** We introduce a *dusting* counterattack mechanism that disrupts the attack by altering the victim's smart contract state. This minimal intervention is timed to invalidate the attack atomicity while providing cost-effective mitigation without replicating attack transactions (i.e., front-running).
- **Stealthy Counterattack:** To prevent front-running risks, FlashGuard bypasses the public mempool and interacts directly with the miners through private relays for stealthy counterattack transactions.
- **EVM Blockchains Applicability:** We show that FlashGuard can be deployed on any EVM-compatible blockchain, such as Ethereum, Arbitrum, Polygon, Avalanche, and Base, to demonstrate its broad generalizability.
- **Empirical Validation:** We evaluate FlashGuard on 20 real-world historical attacks and FlashGuard achieves an average detection latency of 150.31ms with a minimal false positive rate of 0.074%. FlashGuard could have prevented up to $405.71 million in losses.

# 2 BACKGROUND

## 2.1 Transactions on Ethereum

**Ethereum.** Ethereum is a decentralized ledger that supports smart contracts [26]. There are two types of accounts on Ethereum: Externally Owned Accounts (EOAs) and Smart Contract Accounts (SCAs)[45]. In Ethereum Virtual Machine (EVM) based blockchains, EOAs are used to sign sent transactions and are controlled by private keys. SCAs are smart contracts that execute encoded agreement logic. Smart contracts are used to govern on-chain protocols but are also used to craft attacks[46].

**Transactions.** Transactions are used to transfer assets and interact with smart contracts. For each transaction , the nonce ensures each transaction is unique to prevent replay attacks. The gas price influences the transaction's priority to be included in the blockchain and is part of the fee paid to miners. The maximum computational resources that the transaction can use is determined by the gas limit [54].

**Atomic Transactions.** A transaction with more than one internal transaction is an atomic transaction, and if any operation fails, the entire transaction is rolled back [44]. DeFi protocols typically need atomicity in order to reduce risks in executing transactions that have multiple steps. Atomicity is implemented on EVM-based blockchains using the *REVERT* function [37], which undoes all changes when the smart contract's expected conditions are violated, causing the transaction to fail.

**Mempool.** Mempool, or memory pool, temporarily stores all pending transactions [17]. These transactions wait to be included in the blockchain before they are confirmed and included in a block. It allows transaction prioritization based on gas prices as miners typically select transactions with higher fees to be included in the next block to maximize their profit from mining [35]. The mempool is publicly visible, which provides a short window where transactions can be analyzed before confirmation. This short period of time allows detecting and mitigating malicious activities before they are included. For example, this period can be used to detect front-running attacks [30]. Also, the mempool propagates pending transactions , so each node in the network has a copy of the pending transactions until they are confirmed [28].

## 2.2 Flash Loans attacks and On-chain Defenses

**Flash Loans.** Flash loans enable users to borrow large amounts of cryptocurrency without collateral [3]. However, they must be repaid within the same block. If the loan is not repaid in the same block, the transaction is reverted, causing all other internal transactions to revert. It is built on the atomicity property of blockchain transactions to prevent the lender from being exposed to potential risks [41]. Flash loans can be used in different financial operations, such as arbitrage to exploit price discrepancies in DEXs to gain profit [19]. Flash loans also enable swapping of collateral to allow users to change the collateral backing a loan in DeFi protocols without the need to repay them and reinitiate the loan. Additionally, they are used to repay a portion of a loan in self-liquidation to avoid loss of collateral. However, flash loans have also been used in several high-profile attacks, including price manipulation and oracle manipulation [41]. Price manipulation attacks involve using large, borrowed sums to artificially increase or decrease the prices of assets on DEXs, creating arbitrage opportunities. For instance, an attacker might manipulate the price of a token on Uniswap [5] to profit from discrepancies on another platform. On the other hand, oracle manipulation exploits weaknesses in price oracles by feeding incorrect prices to smart contracts. This leads to inaccurate calculations and enables exploits.

**On-chain Defenses.** The idea of deploying on-chain methods to protect DeFi platforms has gained attention due to increasing losses from exploits. These attacks target smart contract vulnerabilities, which can bypass traditional testing methods. This is because no matter how rigorously the smart contract is tested to minimize vulnerability, flaws can still be exploited by new attack vectors, such as flash loan exploits. On-chain defense tools [12, 42] contribute to mitigating flash loan attacks in real-time and provide not only monitoring and defense but also auditing smart contract services. Auditing is a process that evaluates smart contracts' correctness, vulnerabilities, and readiness for deployment, but it does not guarantee the absence of vulnerabilities. Even with these efforts, many audited smart contract vulnerabilities are still exploited and they result in severe losses for DeFi platforms. For example, Euler Finance was exploited by a non-price flash loan attack and lost $197 million [43]. Similarly, in April 2024, Hedgey Finance suffered losses of $44.7 million. These attacks have occurred despite the fact that both of these protocols had undergone full auditing [20, 21, 50]. This presses the need for more robust and proactive on-chain measures tailored to counteract these evolving threats in real-time.

**Table 1:** Limitations solved by FlashGuard (**RTP:** Real-time Prevention, **RTD:** Real-time Detection, **FL:** Flash Loan)

| Study | RTP | RTD | Bypass Mempool | Non-Price FL |
|---|---|---|---|---|
| *FlashGuard* | ✓ | ✓ | ✓ | ✓ |
| FlashSyn [16] | ✗ | ✓ | ✗ | ✗ |
| STING [59] | ✓ | ✗ | ✗ | ✗ |

## 2.3 Limitations of Existing Defense Methods

Existing solutions to combat attacks against smart contract vulnerabilities can be broadly categorized into detection techniques and mitigation methods.

**Detection Techniques.** Existing methods for detecting flash loan attacks are generally not sufficient and have significant limitations (Table 1). Typically, a flash loan transaction is detected by identifying a transaction with borrowed and repaid amounts to the same provider at the end of the transaction. However, the challenge arises in identifying if that flash loan transaction is actually an attack. Previous studies [22] attempt to overcome this challenge by defining a profit threshold for the initiator of the transaction. If the threshold is exceeded, the transaction is classified as an attack. However, this approach can be highly inaccurate, as numerous flash loan transactions yield significant profit, yet they are not attacks. For instance, in arbitrage [53], a user buys a token on a DEX and sells for a higher price on another DEX, exploiting the price difference. With access to large funds, such as flash loans, the profit can be significant but it is not necessarily an attack.

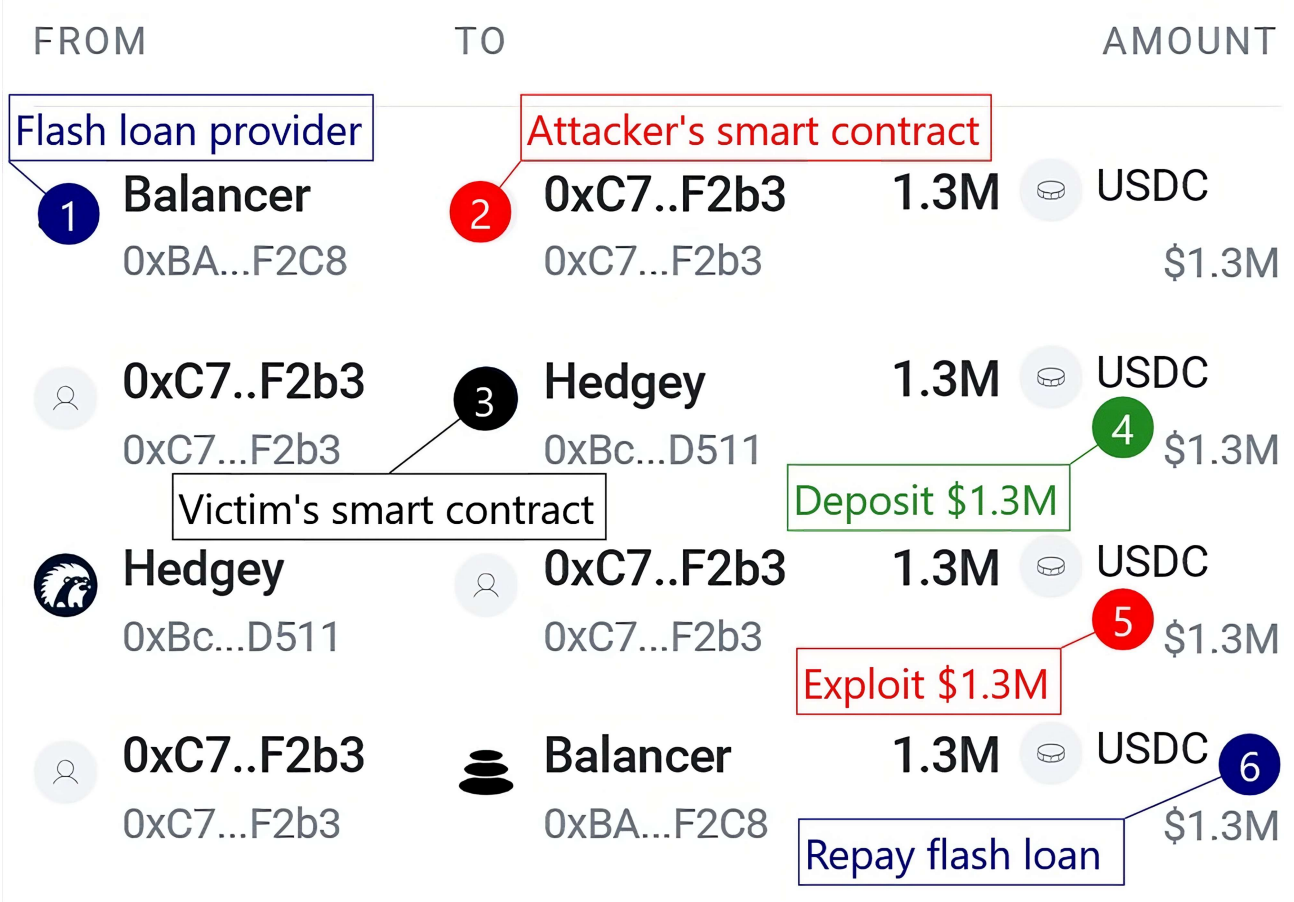

**Figure 1:** Non-price flash loan attack exploited Hedgey Finance.

**Mitigation Methods.** Mitigation in real-time is also a challenging issue. Studies that have tried to overcome this challenge are either focused on non real-time mitigation or they do not provide a detection mechanism. In both cases, this makes the solution less effective as both are essential to stopping the attack before the transaction of the stolen assets is committed to the blockchain. Some studies such as [59] in Table 1 provide only a mitigation solution that is in real-time, but their method is generic for all smart contract attacks and does not provide a detection mechanism, which makes the solution less effective. Also, even though previous solutions include price manipulation attacks, they are not tailored to non-price flash loan attacks. This is critical because such attacks need a prompt response in the limited time window before the execution of the attack is finalized. In Ethereum, for instance, on average, a block is minted every ≈12-13 seconds [32], which adds to the challenge as these attacks have complex interactions and therefore, countermeasures need to be efficient given the limited timeframe.

# 3 NON-PRICE FLASH LOAN ATTACKS

When a flash loan is used to attack a DeFi protocol to exploit a smart contract logic vulnerability, it is classified as a non-price flash loan attack [56]. Non-price flash loan attacks do not depend on price manipulation (i.e., arbitrage)[41], instead, the attacker uses a flash loan in complex smart contract function calls (i.e., misconfigured permissions [33]) to exploit the code in a single atomic transaction. The attacker's goal is to steal funds by exploiting the vulnerability in the smart contract. The stolen assets are typically in ERC20 tokensthat are fungible digital assets on EVM blockchains that comply with the ERC-20 standard [52]. They can be in a single ERC20 token (i.e., USDC) or a combination of tokens (i.e., USDC, MATIC, and WBTC). In both cases, the ownership of ERC20 tokens is transferred to the attacker before it is transferred to the attacker's wallet in the same transaction and before the flash loan is repaid after the attack is finalized[43, 50].

## 3.1 Real-world examples

There are several real-world incidents of non-price flash loan attacks. For instance, Euler Finance, which is a well-known lending protocol on Ethereum was exploited by a non-price flash loan attack on March 13, 2023 [20]. The platform lost $196 million, and the attacker bridged the assets from the BNB Smart Chain (BSC) [11] to Ethereum and used Tornado Cash, a token mixing service, to hide their traces. A similar attack targeted JIMBO, a DeFi platform on Arbitrum, in May 2023 [43]. The attack exploited a logic flaw in its slippage control mechanism, which is a function designed to limit token price fluctuations. This allowed the attacker to steal 4,000 ETH, which was worth $7.5 million at that time. The stolen funds were later moved to Ethereum via the Stargate bridge and the Celer network [23, 40]. Platypus Finance, an Automated Market Maker (AMM) on Avalanche, was also targeted by a flash loan exploit. The attacker minted 41.7 million unauthorized USP tokens through a vulnerability in its liquidity stability control function which resulted in a loss of $8.5 million. While some funds were recovered through Tether (USDT) and Circle (USDC), the attacker successfully bridged $2.4 million USDC out of reach using Gnosis Proxy [29].

We present a case study of the vulnerability in the Hedgey Finance platform that was exploited in a non-price flash loan attack. Hedgey Finance [1] distributes and locks tokens for other DeFi projects. If a DeFi platform intends to distribute tokens (i.e., airdrop to its users), it uses Hedgey Finance to manage the locking and distribution of tokens. However, Hedgey Finance had an unknown vulnerability in creating a campaign function, namely the *createLockedCampaign* function.

Figure 2 shows a simplified view of the Hedgey Finance contract that was exploited and we outline the attack sequence in Figure 1 to show how the flash loan was borrowed, used in the exploit, and repaid. To exploit the vulnerability, the attacker first deployed a smart contract designed for exploiting the Hedgey Finance protocol. The attacker's smart contract address is *0xC7...F2b3*, and the Hedgey exploited smart contract address is *0xBc..D511* as shown in Figure 1. The attacker first took a flash loan from the Balancer[10] vault and targeted the Hedgey Finance smart contract. The attacker's smart contract then called the *createLockedCampaign* function (line 3) to create a campaign using the funds obtained from the flash loan. In this function, once the campaign is created and locked, the user interacting with it receives approval to spend the same amount deposited in case of campaign cancellation (line 12). However, the attacker immediately canceled the campaign by calling *cancelCampaign* on line 15. Then, the funds locked from *createLockedCampaign* were refunded. Nevertheless, there was no revocation of the prior approval in line 12. This allowed the attacker to receive a refund from the canceled campaign and retain approval to spend the same amount from the flash loan. Therefore, the cancellation of the campaign without revoking the approval allowed the attacker to withdraw nearly $2 million from the Hedgey Finance vault [20].

## 3.2 Attack Model

We present a generalized model to capture all non-price flash loan attacks in DeFi platforms. Our model is drawn from real-world

```
1  contract ClaimCampaigns {
2
3      function createLockedCampaign(id, campaign, claimLockup)
           external {
4          require(!usedIds[id], "Campaign in use");
5          require(campaign.amount > 0 && campaign.end > block.
               timestamp, "Invalid campaign");
6
7          usedIds[id] = true;
8          campaigns[id] = campaign;
9          claimLockups[id] = claimLockup;
10
11         // Vulnerability: Approving tokenLocker with campaign
               amount
12         IERC20(campaign.token).approve(claimLockup.tokenLocker
               , campaign.amount);
13     }
14
15     function cancelCampaign(id) external {
16         require(campaigns[id].manager == msg.sender, "Not
               manager");
17
18         delete campaigns[id];
19         delete claimLockups[id];
20
21         // Vulnerability: Allowance to tokenLocker not revoked
22         emit TokensClaimed(id, msg.sender, campaign.amount);
23     }
24 }
```

**Figure 2:** Simplified view of the Hedgey Finance smart contract

attacks, such as the examples presented in Section 3.1 and it provides a framework for understanding these attacks and informs our solution design. Our approach builds upon prior work that characterizes the properties of flash loan exploits [13, 22] but focuses specifically on non-price vulnerabilities.

We assume that the attacker ($A$) and the victim contract ($C$) operate on an EVM-compatible blockchain that supports smart contracts executed by the Ethereum Virtual Machine (EVM), including forks [58] that use the same codebase of Ethereum such as BNB Smart Chain (BSC) [11]. The attacker's transaction (tx) is atomic, meaning all operations succeed or fail together, consisting of multiple internal txs, and passes through the mempool (a temporary storage for unconfirmed transactions). The functions referenced below are deterministic and produce the same output for a given contract state, vulnerability, and input. The following steps detail the sequence of actions for executing a non-price flash loan attack:

① **Acquire a flash loan:** The attacker $A$ borrows a flash loan ($F$) from a flash loan provider ($L$) for use in a single atomic tx $F = \text{Loan}(A, L, \text{amount}, \text{fee})$, where $F > 0$. The repayment is atomic $\text{Repay}(F) \in \text{tx}_{\text{atomic}}$. Here, Loan and Repay are functions provided by $L$ that respectively grant and reclaim borrowed liquidity. The fee is an optional parameter where fee=0 is possible as not all $L$ requires a premium. This is different from the gas fee, which is required by all transactions txs on the blockchain.

② **Satisfy precondition:** The attacker once has $F$, they use it to override conditions required to call the vulnerable function $FC$ in $C$. Typical conditions include token approvals or specific input parameters. In the code example discussed in Section 3.1 example, this is equivalent to *approve* to spend an equal amount of $F$ deposited.

③ **Invoke vulnerable functions:** The attacker invokes a vulnerable function $FC(C, \text{state}, \text{inputs})$ that transitions the contract from state to a corrupted state, state$'$ $FC(C, \text{state}, \text{inputs}) \rightarrow \text{state}'$. Here, state represents the $C$'s internal $FCs$, variables, while inputs are attacker parameters which is $F$.

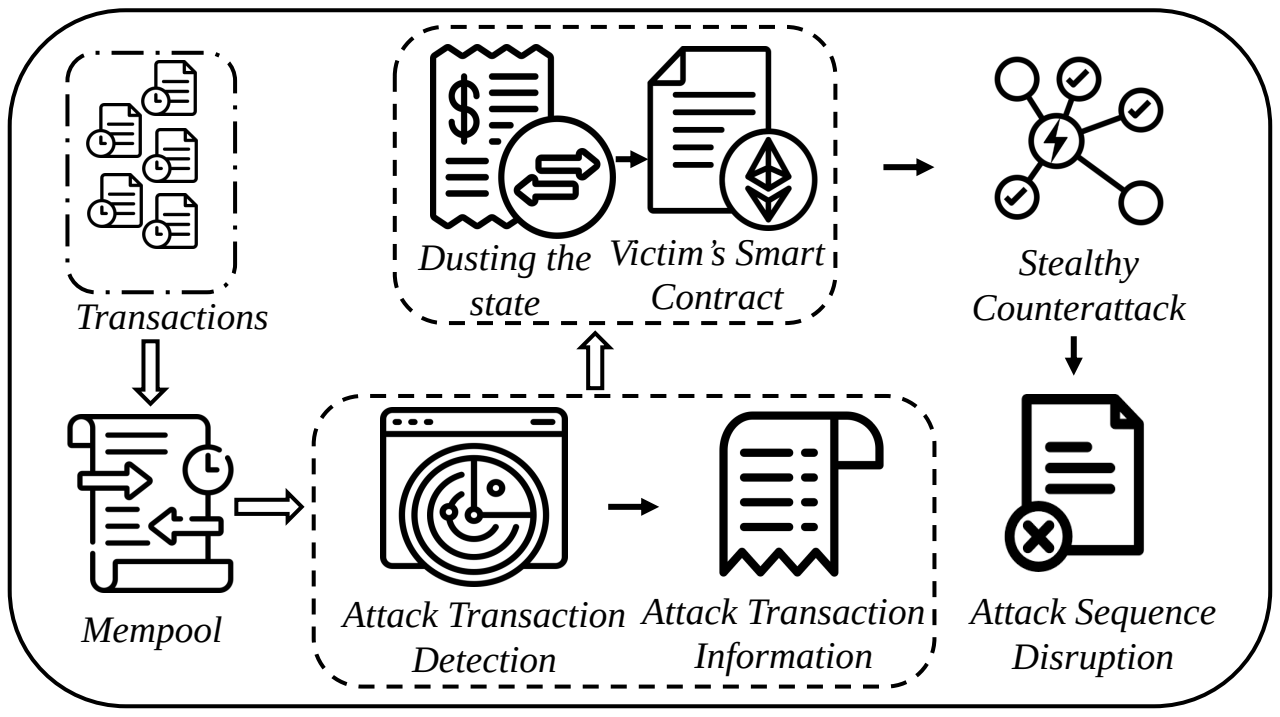

**Figure 3:** FlashGuard overview.

④ **Manipulate contract state:** The attacker changes the state of $C$ by applying $\text{state}'' = \text{CorruptState}(\text{state}', F)$, where CorruptState modifies state′ of the smart contract using $F$. This modification results in unauthorized actions, such as minting assets or preventing token allowance revocation.

⑤ **Overtake assets:** Next, the attacker extracts funds from $C$, $P = \text{assets}(C, \text{state}'', A)$, where $P$ represents ERC20 tokens or stablecoins. This captures any unauthorized transfer to $A$ which represents the exploited assets.

⑥ **Repay the flash loan:** Finally, since tx is atomic, the attacker must repay $F$ to $L$ ($\text{Repay}(F, L)$), and the exploit succeeds only if all steps in $\text{tx}_{\text{atomic}}(\text{Borrow}, \text{Exploit}, \text{Repay})$ are completed. The borrowed amount plus fees, if any, are returned to $L$ in the same transaction.

After the attacker repays $F$, the exploit transaction (tx) is complete, and the attacker has successfully exploited the non-price flash loan vulnerability. We show the main steps of the attack model for the real-world examples from Section 3.1 in Table 2.

**Table 2:** Real-world examples of non-price flash loan attacks

| Attack | Precondition | State Corruption | Overtaken Assets |
|---|---|---|---|
| **Hedgey Finance** | Overriding token approval | Unrevoked token approvals | $2M in stablecoins ($1.3M in one transaction) |
| **Platypus Finance** | Bypassing the USP solvency check | Minting unauthorized tokens | $8.5M in stablecoins |
| **JIMBO** | Exploiting slippage control | Manipulating token swaps | $7.5M in ETH |

## 3.3 Challenges

Detecting and mitigating non-price flash loan attacks involve multiple challenges. Looking at the attack model and the exploited DeFi protocol in Section 3.1, flash loan borrowing (Step 1) and attack initiation (Step 2) in Figure 1 must be detected before exploitation of the funds (Step 5). In addition, any real-time mitigation, such as counterattacks, must be submitted to the blockchain nodes (i.e., miners or validators) before steps 5 and 6 are completed. Therefore, an effective solution to tackle flash loan attacks should address the following challenges:

(C1) **Stopping attacks before they are finalized.** Non-price flash loan attacks operate rapidly in an atomic transaction within a single block, requiring proactive measures to prevent losses. Hence, the defense solution needs to incorporate a real-time strategy to disrupt the attack transaction and protect DeFi assets before it is included in the blockchain.

(C2) **Identifying complex and unique attack patterns.** Non-price flash loan attacks are generally difficult to detect as each attack uses a unique pattern with sophisticated strategies. The absence of collateral for such loans substantially reduces the access barriers of large-scale manipulation operations, which allows anyone to launch a large-scale exploit [13, 22].

(C3) **Protecting DeFi contracts on any EVM blockchains.** Many DeFi protocols deploy the same contracts across multiple blockchains, most of which are EVM-compatible. If a vulnerability is identified on one blockchain, it is trivial for attackers to exploit the same vulnerability on another. For example, the attack that hit Hedgey Finance caused nearly $44 million [20] in losses for the same protocol running on two different blockchains.

(C4) **Preventing irrecoverable losses.** In most cases, non-price flash loan attacks result in significant financial losses as DeFi protocols manage substantial asset vaults. Mitigating these attacks needs timely countermeasures because once the attack is finalized and the transaction is confirmed, the lost funds cannot be recovered involuntarily. This is evident from the fact that only about 8.3% of the total assets lost in blockchain attacks have been recovered [20].

# 4 FLASHGUARD

FlashGuard is designed to address the key challenges of detecting and mitigating non-price flash loan attacks in DeFi platforms. At a high level, FlashGuard consists of three main components (Figure 3) as described below:

**Event-based Attack Detection.** FlashGuard uses real-time analysis of pending transactions function signatures (unique function identifiers for smart contract methods) in the mempool to detect non-price flash loan attacks. This identifies attack transactions that exploit smart contract logic vulnerabilities using flash loans.

**Dusting Counterattack.** When the attack is detected, FlashGuard passes the attack details to the disruptor module. This module initiates a *dusting* transaction, which is a small ERC20 token transfer to or from the victim's smart contract, after the attacker calls the victim's smart contract but before the attack confirmation. It modifies the contract's state and disrupts the atomicity of the attacker's transaction. This counterattack invalidates the exploit.

**Counterattacks using Private Relays.** FlashGuard submits the *dusting* transaction through a private relay [25] which prevents it from appearing in the public mempool while ensuring that the counterattack is confirmed before the attacker's transaction.

## 4.1 Event-based Attack Detection.

Attack detection in FlashGuard aims to identify suspicious transaction hashes for the disruptor to prevent the attack from being finalized by monitoring the mempool in real-time, which is a temporary storage for pending transactions. It identifies potential non-price flash loans that exploit smart contract logic before attack transactions are minted on the blockchain.

**Attack Transactions.** We observe that the attack transactions

**Algorithm 1** FlashGuard Mempool Detection

**Require:** $\mathcal{T}$: Mempool transactions, $\mathcal{E}$: Excluded addresses, $\mathcal{S}$: Function signatures
**Ensure:** $\mathcal{Q}$: Set of attack transactions

$\mathcal{Q} \leftarrow \emptyset$
**for all** $tx \in \mathcal{T}$ **do**
    $from_addr \leftarrow tx$.from
    $to_addr \leftarrow tx$.to
    **if** $from_addr \in \mathcal{E}$ **or** $to_addr \in \mathcal{E}$ **then**
        **continue**
    **end if**
    $input_data \leftarrow tx$.input_data
    $functions \leftarrow$ DecodeInputData($input_data$)
    $flashLoan \leftarrow$ False
    $transfer \leftarrow$ False
    $withdrawal \leftarrow$ False
    $approval \leftarrow$ False
    **for all** $func \in functions$ **do**
        **if** $func$.signature $\in \mathcal{S}_{\text{flashLoan}}$ **then**
            $flashLoan \leftarrow$ True
        **else if** $func$.signature $\in \mathcal{S}_{\text{transfer}}$ **then**
            $transfer \leftarrow$ True
        **else if** $func$.signature $\in \mathcal{S}_{\text{withdrawal}}$ **then**
            $withdrawal \leftarrow$ True
        **else if** $func$.signature $\in \mathcal{S}_{\text{approval}}$ **then**
            $approval \leftarrow$ True
        **end if**
    **end for**
    **if** $flashLoan$ **and** ($transfer$ **or** $withdrawal$) **and** $approval$ **then**
        $\mathcal{Q} \leftarrow \mathcal{Q} \cup \{tx.\text{hash}\}$
    **end if**
**end for**
**return** $\mathcal{Q}$

typically share a combination of function signatures used for identifying specific methods. They are usually emitted when the transaction calls the smart contract and can be captured from the transaction input data in the mempool before the transaction confirmation. Therefore, the detection targets a combination of these signatures as a signal of an attack initiation. The signatures $\mathcal{S}$ include *FlashLoan*, token *transfers*, *approvals*, and *withdrawals* to a smart contract or wallet. For a transaction $tx$, its decoded function signatures are $\mathcal{F}(tx)$ = DecodeInputData($tx$.input_data). The input data, $tx$.input_data, encodes data of the smart contract function being called and its parameters. These parameters represent the values passed to the called function. However, parameter value changes do not change the function signature, but functions related specifically to flash loans from different providers can have distinct signatures. The *FlashLoan* signature, for instance, can differ and depend on the flash loan provider.

Each flash loan provider may have a distinct event signature for the function used to lend a flash loan when interacting with the blockchain. Table 3 shows the function signatures for different flash loan providers and widely used functions such as *transfers*, *approvals*, and *withdrawals* have the same signatures across EVM-blockchains. Each signature is a 256-bit (32-byte) value [2].

A combination of these signatures is in fact a signal of a non-price flash loan attack. A transaction $tx$ consists of a flash loan if $\mathcal{F}(tx) \cap \mathcal{S}_{\text{flashLoan}} \neq \emptyset$, where $\mathcal{S}_{\text{flashLoan}}$ denotes the set of flash loan providers' function signatures. Similarly, $tx$ is an attack if it contains transfers, approvals, or withdrawals such that $\mathcal{F}(tx) \cap \mathcal{S}_{\text{transfer}} \neq \emptyset$, $\mathcal{F}(tx) \cap \mathcal{S}_{\text{approval}} \neq \emptyset$, or $\mathcal{F}(tx) \cap \mathcal{S}_{\text{withdrawal}} \neq \emptyset$, respectively.

These function signatures may overlap with known benign addresses, such as MEV bots that search and front-run profitable transactions.[19]. This is especially true for those transactions when an MEV bot spots a profitable opportunity. In such cases, it initiates a flash loan transaction to secure it, but not all MEV bots are designed to borrow flash loans and therefore, FlashGuard carefully excludes these known addresses in its detection approach. We discuss address exclusion further in the detection approach.

**Table 3:** Examples of function signatures used for the mempool attack detection.

| Function | Function Signature | Event Signature |
|---|---|---|
| `flashLoan(address,uint256,...)` | `0x1b8b5af1` | `0x0d7d7...95f0` |
| `flashLoan(address,uint256,...)` | `0xc2b12a73` | `0xf4626...22ab` |
| `transfer(address,uint256)` | `0xa9059cbb` | `0xddf25...b3ef` |
| `approve(address,uint256)` | `0x095ea7b3` | `0x8c5be...b925` |
| `withdraw(uint256)` | `0x2e1a7d4d` | `0xe1fff...39db` |

**Note: Function signatures are 4-byte identifiers derived from the first four bytes of the Keccak-256 hash of the function.**

**Attack Identification** Identifying the above-mentioned events in a single transaction signifies an attack transaction as a non-price flash loan attack relies on executing the attack in an atomic transaction. These events manifest as internal transactions and interact with different smart contracts as atomic transactions. This increases the transaction complexity, making it challenging to decode it in a timely manner before the transaction is included by the miners [9]. Once the transaction is included in the blockchain, the stolen assets belong to the attackers. Note that reducing irrelevant transactions and focusing on only possible attack transactions is important to make the detection run faster [47].

**Detection Approach and Detection Parameters.** To identify and mitigate attacks on any EVM blockchain, FlashGuard employs an EVM-compatible approach for real-time mempool monitoring, shown in Algorithm 1. A transaction $tx \in \mathcal{T}$, where $\mathcal{T}$ represents all mempool transactions, is excluded if the sender ($tx$.from) or receiver ($tx$.to) of $tx$ is on the exclusion list $\mathcal{E}$, which is a list of known benign addresses compiled based on historical on-chain interactions with smart contracts to reduce misclassification of attacks. As we have mentioned, some MEV bots are benign with the goal of gaining profit by front-running transactions. However, in some cases, they front-run attack transactions. In this case, the MEV bot returns the funds to the saved smart contract and deems it as a rescue transaction [36]. Therefore, $tx.\text{from} \in \mathcal{E} \lor tx.\text{to} \in \mathcal{E}$. $tx$ is scrutinized if it satisfies $\mathcal{F}(tx) \cap \mathcal{S}_{\text{flashLoan}} \neq \emptyset \land \mathcal{F}(tx) \cap (\mathcal{S}_{\text{transfer}} \cup \mathcal{S}_{\text{withdrawal}}) \neq \emptyset \land \mathcal{F}(tx) \cap \mathcal{S}_{\text{approval}} \neq \emptyset$. This means that all attack transactions call *flashLoan* and *approve* to spend exploited tokens but could interact with either *transfers* or *withdraw* or both to obtain assets depending on the target smart contract's vulnerability.

Any $tx$ that does not include these function signatures is removed from being a potential attack, which makes the detection more efficient. In general, accurate detection depends on identifying unique patterns of flash loan transactions. These are based on function signatures corresponding to known flash loan providers such as Balancer and dYdX smart contracts' function calls. These signatures serve as markers within the transaction events. The transactions flagged as an attack are denoted as $\mathcal{Q} = \{tx.\text{hash} \mid tx \in \mathcal{T}, \neg(tx.\text{from} \in \mathcal{E} \lor tx.\text{to} \in \mathcal{E}), \text{Qualify}(tx)\}$. Once the attack is detected (Algorithm 1), the transaction hash is dispatched to the *counterattack* component of FlashGuard to stop the attack.

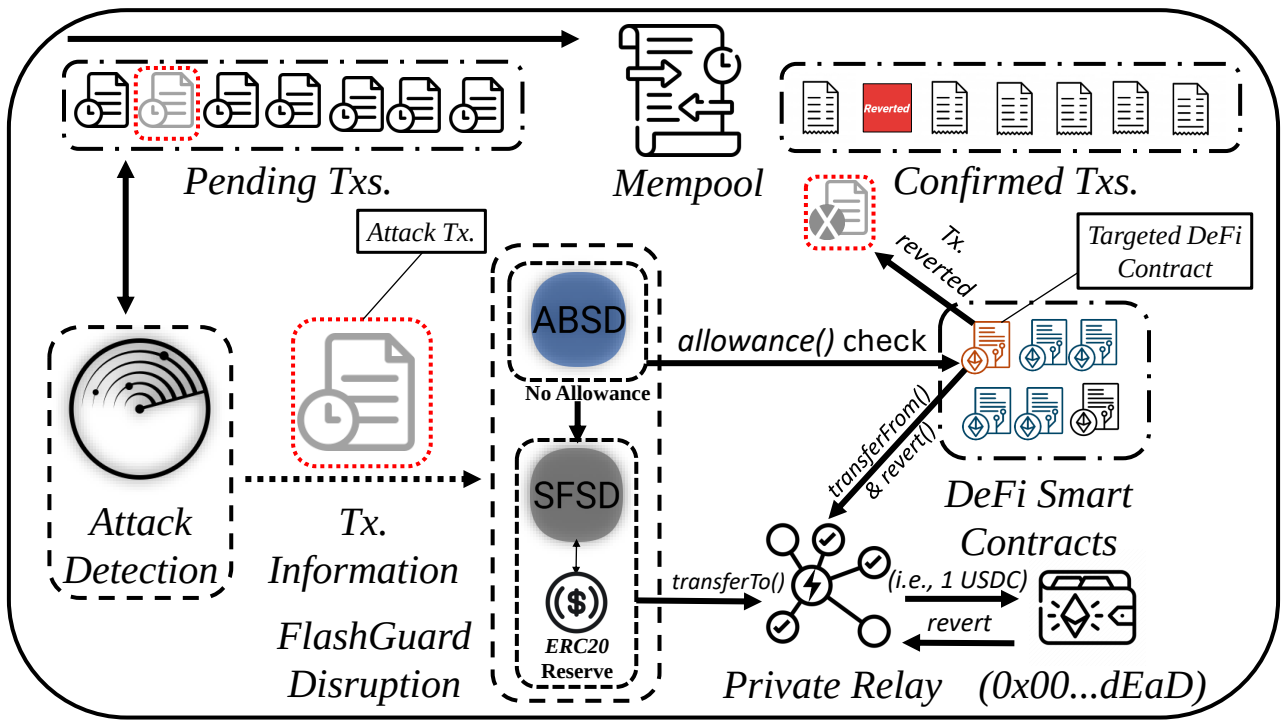

**Figure 4:** *Dusting* counterattack showing ABSD and SFSD mechanisms.

## 4.2 Dusting Counterattack

The *dusting* process in FlashGuard initiates a counterattack transaction that changes the current state of the victim smart contract after attack detection. This occurs after the attack transaction calls the victim's vulnerable smart contract. As discussed in Section 4.1, when FlashGuard detects a potential attack, the set of attack transactions $\mathcal{Q}$ identified in the detection proceeds to the disruption and FlashGuard immediately dispatches a *dusting* counter-transaction. *dusting* can be performed in two ways as discussed next.

**Allowance-Based State Dusting (ABSD).** DeFi platforms can subscribe to FlashGuard to counterattack potential non-price flash loan attacks. Once subscribed, FlashGuard will be granted a small allowance `allowance` (i.e., 1 USDC) to spend from the DeFi smart contract. This approach reduces reliance on FlashGuard's reserves and increases the chance of a successful counterattack, as it targets the specific smart contract address of the vulnerable function. Here, subscription is more efficient as FlashGuard needs to only call the DeFi-specific disrupter smart contract. Even though the required funds are minimal, FlashGuard needs this allowance to guarantee state *dusting* in case of an attack. FlashGuard determines the token that is involved in the attack by inspecting the `tokenAddress` parameter of the transaction. It then populates the *dusting* amount as dustingAmount = min(`allowance`, fixed_amount), where `allowance` is the maximum amount granted to FlashGuard, and fixed_amount is predefined for *dusting*. The current allowance is checked using `IERC20.allowance()`, which verifies how much the vulnerable contract has authorized FlashGuard to transfer. If `allowance` < dustingAmount, FlashGuard terminates the disruption and returns `Failure`. Otherwise, FlashGuard uses `IERC20.transferFrom()` to transfer dustingAmount from the vulnerable contract to a burn address[1]. This alters the expected state of the vulnerable contract after the attack transaction call. FlashGuard then calls `revert()` to invalidate the attack transaction and return the dusted value to the DeFi smart contract. Figure 5 considers a DeFi platform where FlashGuard has been granted an allowance of 2 USDC. Suppose a flash loan attack is detected to target this platform, FlashGuard uses the `IERC20.transferFrom()` function to transfer 1 USDC (dustingAmount) from the DeFi smart contract to the burn address and then calls `revert()` after the attacker's call. This alters the contract's state between the time of the attack call and the attempt to confirm the transaction. The attacker's transaction will no longer be valid as its prior assumption about the DeFi smart contract is invalidated by changing its state, which triggers the EVM to revert to preserve atomicity and undo all internal transactions.

[1](`0x000000000000000000000000000000000000dEaD`)

**Algorithm 2** FlashGuard Disruption

**Require:** $tx$: Detected transaction, $\mathcal{S}$: Function signatures
**Ensure:** Attack transaction is disrupted

$to_addr \leftarrow tx$.to
$token \leftarrow$ GetToken($tx$)
$dustingAmount \leftarrow$ CalculateDustingAmount($token$)
**if** isSubscribed($to_addr$) **then**
  $allowance \leftarrow$ GetAllowance($to_addr, token$)
  **if** $allowance < dustingAmount$ **then**
    **return Failure: "Insufficient allowance for disruption"**
  **end if**
  TransferFrom($to_addr$, BurnAddress, $dustingAmount$)
**else**
  Transfer($token$, BurnAddress, $dustingAmount$)
**end if**
**Revert transaction:** "Disruption successful"

**Self-Funded State Dusting (SFSD).** If the DeFi platform has not subscribed to FlashGuard, no allowance is granted to use the platform's tokens. In this case, FlashGuard disrupts the attack using its own funds. Note that the fund used, even though it is small, will be returned to FlashGuard as a result of the `revert`. However, FlashGuard will minimize the gas fee spent to counter attacks under this model, which may result in attack evasion for non-ABSD DeFi smart contracts. When a malicious transaction $tx \in \mathcal{Q}$ is detected, FlashGuard assigns dustingAmount = fixed_amount and initializes the token interface using `IERC20(tokenAddress)`. FlashGuard directly calls `IERC20.transfer()` to send dustingAmount from its own reserves to the burn address. This changes the state of the vulnerable contract and invalidates the attack transaction's assumption about the atomicity. Finally, FlashGuard calls `revert()`, which forces the attack transaction to fail and prevents the attacker from completing their exploit. FlashGuard attack disruption for these two scenarios under ABSD and SFSD are shown in Figure 5.

## 4.3 Counterattacks using Private Relays

Both ABSD and SFSD disrupt non-price flash loan attacks by introducing a minimal state change in the victim smart contract. We note that there is no specific dustingAmount required in this process, as only the minimal value is transferred and then reverted using `revert()`, causing the attack transaction to fail. However, the most effective disruption occurs under the ABSD model when *dusting* is executed after the attacker borrows the flash loan and calls the victim smart contract before the attack transaction is confirmed. If dusting occurs before the attacker calls the victim contract or after the confirmation, it has no effect. Hence, timing is critically important. To address this, FlashGuard utilizes private relays [25]. Transactions submitted through private relay bypass the public mempool and are delivered directly to miners. This provides a low-latency confirmation time for the *dusting* transaction and avoids the delays incurred during the public mempool transaction confirmation.

Most transactions get added to the blockchain via the mempool [49]. In the mempool, transactions are queued, waiting for the miners to include them in the next block and they are prioritized based on gas fee. Our approach includes the *counterattack* transaction before the attack transaction is committed to the blockchain.

**Table 4:** Non-price flash loan attacks detection computational overhead in DeFi for various blockchain networks, Ethereum (ETH), Base (Base), Arbitrum (ARBI), Polygon (Poly), and Avalanche (AVAX). PMU stands for Peak Memory Usage. **DeT** is the detection time.

| Attack | Date | Chains | Loss | Detected | DeT(ms.) | Avg. CPU(%) | Avg. PMU(MiB) |
|---|---|---|---|---|---|---|---|
| Hedgey Finance | 04-19-2024 | ETH | $2.00M | ✓ | 129.93 | 1.80 | 0.53 |
| Sumer Money | 12-04-2024 | Base | $350K | ✓ | 122.93 | 4.10 | 0.45 |
| LavaLending | 03-29-2024 | ARBI | $340K | ✓ | 148.91 | 3.46 | 0.87 |
| PRISMAFI | 03-28-2024 | ETH | $11.60M | ✓ | 142.92 | 3.78 | 0.69 |
| Rosa Finance | 01-18-2024 | ARBI | $44.67K | ✓ | 121.93 | 3.28 | 0.92 |
| Themis Protocol | 06-27-2023 | ARBI | $367.75K | ✓ | 127.93 | 3.04 | 0.17 |
| JIMBO | 05-29-2023 | ARBI | $7.50M | ✓ | 294.83 | 3.25 | 0.57 |
| EON | 06-02-2023 | POLY | $29.20K | ✓ | 110.94 | 35.72 | 0.05 |
| Ovix | 04-28-2023 | POLY | $2.00M | ✓ | 403.77 | 27.82 | 0.70 |
| Euler Finance | 03-13-2023 | ETH | $196.00M | ✓ | 146.42 | 7.44 | 0.10 |
| Platypus Finance | 02-16-2023 | AVAX | $8.50M | ✓ | 76.96 | 8.87 | 0.15 |
| Midas Capital | 01-15-2023 | POLY | $650K | ✓ | 147.91 | 15.57 | 0.72 |
| Cauldron | 09-06-2022 | AVAX | $370K | ✓ | 86.95 | 6.63 | 0.71 |
| Cream Finance | 10-27-2021 | ETH | $130.00M | ✓ | 162.91 | 3.08 | 0.53 |
| XTOKEN | 05-12-2021 | ETH | $24.50M | ✓ | 183.40 | 2.59 | 0.06 |
| Warp Finance | 12-18-2020 | ETH | $7.80M | ✓ | 97.94 | 5.21 | 0.07 |
| Akropolis | 12-11-2020 | ETH | $2.00M | ✓ | 135.92 | 8.15 | 0.16 |
| Origin Protocol | 11-17-2020 | ETH | $8.00M | ✓ | 137.92 | 8.30 | 0.15 |
| Cheese Bank | 11-06-2020 | ETH | $3.30M | ✓ | 111.94 | 5.49 | 0.08 |
| bZx | 02-18-2020 | ETH | $355.88K | ✓ | 113.93 | 7.16 | 0.41 |

```
contract FlashGuardDisruptor {
    function disruptAttack(bool isSubscribed,
        address tokenAddress) external {
        uint256 dustingAmount = 0.1; // Minimal amount "dust" to cause state change
        IERC20 targetToken = IERC20(tokenAddress); // Determine the the token involved in the attack

        if (isSubscribed) {
            // Case 1: DeFi platform subscribed to FlashGuard
            uint256 allowance = targetToken.allowance(
                VULNERABLE_CONTRACT, address(this));
            require(allowance >= dustingAmount, "Insufficient allowance");
            targetToken.transferFrom(VULNERABLE_CONTRACT,
                address(0xdead), dustingAmount);
        } else {
            // Case 2: DeFi platform not subscribed to FlashGuard
            targetToken.transfer(address(0xdead),
                dustingAmount); // Use FlashGuard's funds
        }
        // Revert to disrupt the attack's atomicity
        revert("Disruption successful: Attack invalidated");
    }
}
```

**Figure 5:** The disruption mechanism of FlashGuard handles two cases ABDS and SFDS, whether the DeFi platform subscribes to FlashGuard with a minimal allowance or not. The used fund in the *dusting* transaction is too small to alter the vulnerable contract's state. The token used for *dusting* (i.e., USDC) is determined by the vulnerability the attack attempts to exploit.

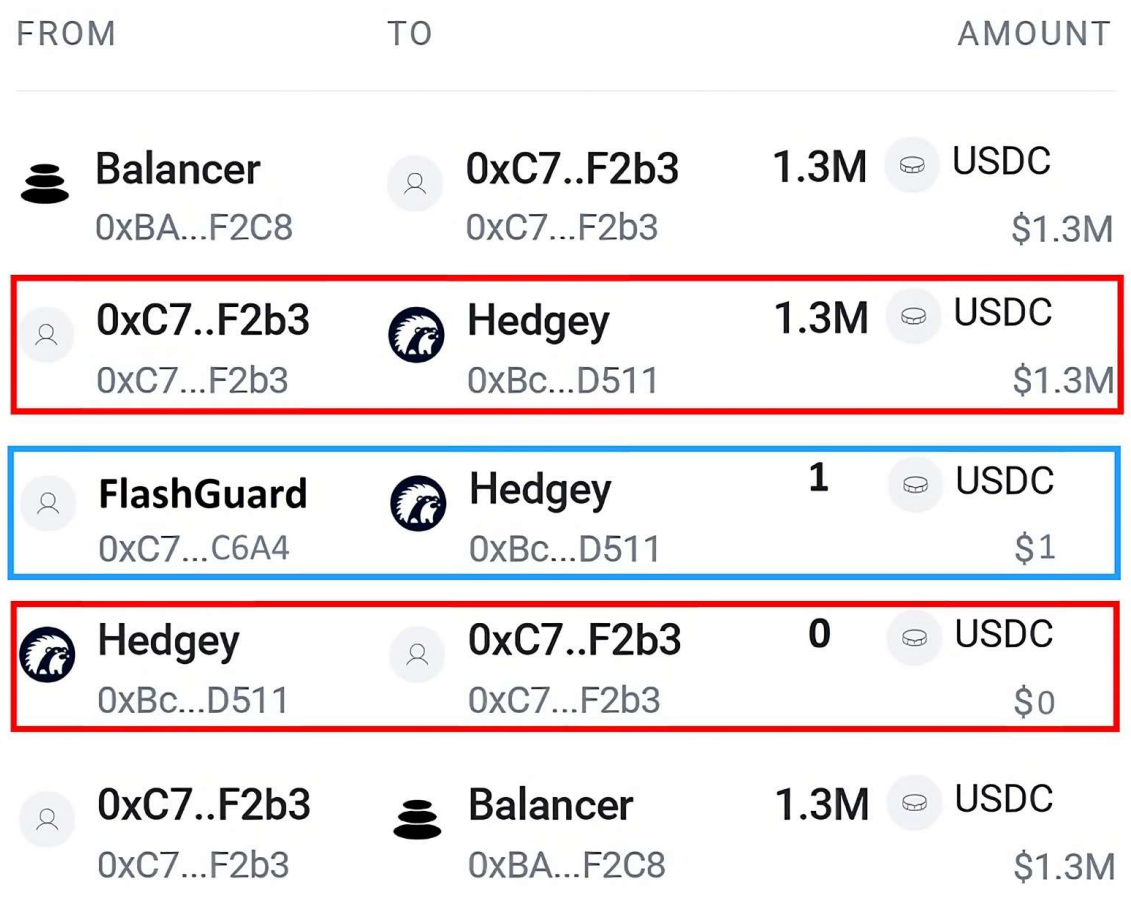

**Figure 6:** Disruption example of the Hedgey Finance exploit based on Figure 1. The figure illustrates the *dusting* counterattack by FlashGuard (in the blue box), which occurs after the flash loan borrowing and forces the attack transactions (in red boxes) to revert. As a result, the flash loan is repaid, and the attacker contract receives 0 USDC.

Therefore, we utilize private relays from *Flashbots* that mitigate the negative effect of MEV on blockchain fairness. MEVs can profit from reordering or inserting transactions and here, blockchain fairness refers to the unbiased transaction inclusion and ordering without miner manipulation such as front-running [25].

MEVs benefit from rearranging transactions in a block to maximize their profit [38]. Thus, private relays provide the ability to submit transactions directly to the blockchain miners which is usually faster but could incur a slightly higher gas fee. For our approach, it provides two benefits. First, it hides the counterattack transaction of FlashGuard from being visible in the mempool and second, it ensures that the transaction is confirmed faster since it does not need to wait in the mempool to be included by miners in the next block.

## 5 EVALUATION

FlashGuard is implemented on top of the Web3[18] using Solidity [48], and Forge blockchain development toolkit [27]. We leveraged Web3's *WebSocket* protocol to monitor the mempool in real time for detecting attack transactions. The real time detection is integrated with multi-chain node providers such as Alchemy [7] and Infura [31]. As FlashGuard provides a generic solution to any evm-blockchain, we evaluate FlashGuard using attacks from different blockchains, namely Ethereum (ETH), Base (Base), Arbitrum (ARBI), Polygon (POLY), and Avalanche (AVAX) as shown in Table 4. The transaction data used in our evaluation are compiled from different sources [20, 43, 50], and they include a large number of DeFi attacks. In addition, we have replicated several unseen attacks for different vulnerabilities from [51] and our study includes a total of 20 real-world non-price flash loan attacks that occurred between 2020 and 2024, in addition to the unseen attacks.

**Table 5:** Comparison with previous studies on detection and mitigation capabilities of real-world non-price flash loan attacks. **DeT** denotes detection and **M** mitigation methods.

| Study | # of Attacks | DeT. | M. | Real-time |
|---|---|---|---|---|
| **FlashGuard** | 20 | ✓ | ✓ | ✓ |
| STING[59] | 2 | | ✓ | ✓ |
| LeiShen[56] | 4 | ✓ | | |
| DeFiRanger[55] | 3 | ✓ | ✓ | |

## 5.1 Attack Replication

We replicate attacks by first modeling the DeFi protocol and then we reproduce the attack transaction, the liquidity pools, and token smart contracts. Since we need to mirror the same state of the blockchain at the time of the attack, we used The Remote Blockchain Call (RBC) [24] of each blockchain for accurately reproducing the DeFi protocol and the attacker's smart contracts with the same state as when the attack occurred. The attack sequence includes the steps for flash loan initiation, repayment as well as the DeFi protocol exploitation. We developed a mock ERC20 token contract called MockUSDC, to reproduce the stablecoin behavior within our test environment. The mock tokens are used to reproduce attacks that had no pre-existing code by controlling the interactions exactly as needed to mimic each attack. Attack replication is recreating the sequence of the attack steps based on the transaction that exploits the DeFi protocol.

For each attack, we granted the allowance and permission needed to interact with the smart contract. The allowance represents the maximum asset amount that an external EAO can spend while interacting with the smart contract. We also preloaded the attacker's address with the necessary balances to carry out the reproduced exploit. The malicious smart contract consists of the exploitation sequence and it is invoked by the attacker. We model the attack sequence, such as flash loans, token swaps, and liquidity manipulation to mimic how the borrowed tokens can be used to exploit the DeFi protocol and steal assets. After each replication, we calculated the attack's value of exploited assets. Note that some of these assets are a mix of tokens. For example, in the exploit of Rosa Finance [20], the attack took DAI, USDC, and WBTC after repaying the flash loan. As shown in Tables 4 and 7 in Section5.3, FlashGuard is able to detect and disrupt all historic attacks. However, to evaluate the generalizability of FlashGuard, we analyze the effectiveness of FlashGuard for new attacks based on the DeFi smart contract vulnerabilities outlined in *DeFiVulnLabs* [51]. These vulnerabilities are not exploited in the historical non-price flash loan attacks but they are potential vulnerabilities that could be exploited at any point in time. For a vulnerability to be realistically exploitable using a flash loan in non-price flash loan attacks, the attack must result in a profit for the attacker. Therefore, we examine relevant vulnerabilities shown in Table 8.

The table shows susceptible DeFi smart contract vulnerabilities that can be exploited by flash loans in non-price flash loan attacks. For example, unchecked return values can allow token transfers to an attacker when the tokens return false instead of reverting. In this case, flash loans would provide large liquidity. Also, improper delegatecall enables attackers to execute arbitrary code in the caller's context, and flash loans can amplify the impact by manipulating large amounts of liquidity. Uninitialized storage pointers could create opportunities for state corruption by exploiting uninitialized variables. Flash loans, in this context, provide the funds needed to leverage this vulnerability. Finally, improper access control allows unauthorized access to restricted functions [59] and flash loans can amplify manipulation by exposing critical operations to unauthorized EOAs, especially when they are token-related.

We reproduce the vulnerabilities shown in Table 8 on the DeFi platform to realize the unseen attacks. We set $50,000 in a stablecoin as the vulnerable funds to a flash loan attack. For each unseen attack, we first observe the expected behavior for the vulnerable function before applying FlashGuard to ensure that the original DeFi function runs with no logical error. In total, we conducted 6 tests for the 3 vulnerabilities consisting of two tests per vulnerability for *ABSD* and *SFDS* respectively.

## 5.2 Metrics

**Detection.** We measure the effectiveness of FlashGuard's detection in terms of detection latency, CPU usage, and memory consumption for each historical attack (Table 4). Detection latency represents the time duration between the occurrence of the transaction in the mempool to its identification as a threat. Achieving low detection latency is important for real-time detection in FlashGuard as even small delays can prevent FlashGuard from stopping an attack. Low CPU and memory usage ensures that FlashGuard operates with minimal overhead and can be integrated into environments with limited computational resources.

**Disruption.** The effectiveness of FlashGuard's disruption is measured using multiple metrics that capture its performance for different attacks in various blockchains. Time(ms) measures how quickly FlashGuard's execution acts from *dusting* to forcing the attack to fail. The total number of transactions (#txs) involved in the attack reflects the attack's complexity. However, not all internal txs are malicious. Some txs are used to set up the targeted exploit. #Entities represents the count of the unique addresses involved in the attack which includes attackers, victim contracts, and intermediaries such as the flash loan provider. DD (Data Dependencies) captures the number of events including approvals and allowances that attackers need to execute their attack. We use MF (Money Flow Events) to track how funds move during the attack. Usually, the exploited assets are transferred to the attacker's authorized wallet or smart contract. TK (ERC20 Tokens) represents the number of tokens affected by the attack. The Gas cost of the counterattack transaction to stop the attack and Price (USD) quantify the financial cost of FlashGuard.

**Table 6:** Detection accuracy from different EVM blockchains for non-price flash loan attacks in our study.

| Blockchain | Transactions | False Positive Rate (FPR) | Accuracy (%) |
|---|---|---|---|
| Ethereum | 475,992 | 0.03% | 99.97% |
| Base | 53,268 | 0.05% | 99.95% |
| Arbitrum | 8,576 | 0.09% | 99.91% |
| Polygon | 68,909 | 0.16% | 99.84% |
| Avalanche | 7,884 | 0.04% | 99.96% |

## 5.3 Experiment Results

**Attack Detection.** FlashGuard successfully detects all the 20 real-world historical attacks (Table 4). Our detection latency shows an average detection time of 150.31ms. We also examine FlashGuard's performance in terms of false positives, where transactions are incorrectly identified as attacks. FlashGuard incurs a low false positive rate of 0.074% of examined transactions. For each identified attack, we analyze 200 blocks before and 200 blocks after the attack (Table 6). If a transaction is false positive but normal, FlashGuard will prevent it from being executed by applying the dusting mechanism described in Section 4.2, which, in the worst case, will cause the transaction to fail. While we found minimal misclassification, shown in Table 6, they can be rectified by excluding the found known normal transactions, such as Maximal Extractable Value (MEV) bots that search for profit using flash loans. We also found some MEV bot transactions flagged as attacks, which we excluded and considered them as non-attacks in our experiments.

**Table 7:** Counterattack disruption metrics from different blockchain networks. **#txs** is the number of transactions, **#Entities** shows the number of entities involved, **DD** is for the number of distinct data dependency events (i.e., approve token), **MF** is money flow events (unique interactions), and **TK** is the count of ERC20 tokens involved in the attack.

| Attack | Chain | Root Cause | Time (ms) | #txs | #Entities | DD | MF | TK | Gas Cost (Gwei) | Cost (Native Token) | Price (USD) |
|---|---|---|---|---|---|---|---|---|---|---|---|
| Hedgey Finance | ETH | Abusing createLockedCampaign | 23.21 | 10 | 3 | 3 | 4 | 1 | 6,933,637.62 | 0.0069 ETH | 16.81 |
| Sumer Money | Base | Manipulating wrapper contracts | 55.06 | 65 | 8 | 1 | 19 | 5 | 126,897 | 0.0038 ETH | 6.09 |
| LavaLending | ARBI | Exploiting lending protocol | 41.48 | 187 | 24 | 16 | 80 | 9 | 32,943.3 | 0.00003 ETH | 0.0593 |
| PRISMAFI | ETH | Logic flaw in migration contract | 7.11 s | 75 | 22 | 4 | 29 | 9 | 51,967,530 | 0.052 ETH | 90.94 |
| Rosa Finance | ARBI | Liquidity index logic vulnerability | 73.36 | 51 | 23 | 3 | 16 | 15 | 7,929.3 | 0.0079 ETH | 0.01957 |
| Themis | ARBI | Vulnerabilities in multiple pools | 64.48 | 129 | 24 | 15 | 46 | 21 | 6,250,187 | 0.0063 ETH | 10.62 |
| JIMBO | ARBI | slippage control in the *shift()* function | 134.88 | 373 | 12 | 16 | 117 | 3 | 107,046 | 0.000107046 ETH | 0.1926828 |
| EON | POLY | Inadequate pool control | 86.97 | 23 | 31 | 9 | 10 | 20 | 2,629,440 | 0.00262944 MATIC | 0.0011219 |
| 0vix | POLY | Exploit in deflationary token contract | 65.4 | 736 | 40 | 66 | 278 | 21 | 948,328.356 | 0.0009 MATIC | 0.000384 |
| Euler Finance | ETH | Improper liquidation checks | 83.41 | 56 | 4 | 2 | 20 | 4 | 2,873,970 | 0.0029 ETH | 4.60 |
| Platypus Finance | AVAX | Issue with LP-USDC conversion | 163.33 | 72 | 17 | 2 | 22 | 19 | 167,375 | 0.0047 AVAX | 0.1232 |
| Midas Capital | POLY | Collateral logic vulnerability | 30.42 | 207 | 33 | 27 | 87 | 3 | 1,149,659 | 0.0011 MATIC | 0.0004656 |
| Cauldron | AVAX | Exchange rate logic flaw | 74.91 | 70 | 15 | 10 | 29 | 8 | 1,210,825 | 0.0012 AVAX | 0.03182 |
| Cream Finance | ETH | Uncapped supply manipulation | 42.33 | 164 | 39 | 5 | 69 | 29 | 13,577,264 | 0.0136 ETH | 21.72 |
| XTOKEN | ETH | Inadequate validation of input | 69.45 | 248 | 18 | 23 | 40 | 3 | 6,221.544 | 0.0062 ETH | 11.20 |
| Warp Finance | ETH | Collateral valuation logic | 25.88 | 57 | 3 | 3 | 12 | 3 | 2,588.439 | 0.0026 ETH | 4.66 |
| Akropolis | ETH | Reentrancy via fake token deposit | 36.40 | 123 | 10 | 8 | 69 | 4 | 3,262.749 | 0.0033 ETH | 5.87 |
| Origin Protocol | ETH | Origin Dollar (OUSD) reentrancy vulnerability | 16.35 | 156 | 3 | 4 | 12 | 2 | 2,877.504 | 0.0029 ETH | 5.18 |
| Cheese Bank | ETH | Collateral valuation logic | 5.28 | 64 | 11 | 21 | 5 | 6 | 6,352.731 | 0.0064 ETH | 11.43 |
| bZx | ETH | Logic flaw in collateral validation | 15.89 | 43 | 14 | 1 | 17 | 6 | 6,942.480 | 0.0069 ETH | 12.49 |

FlashGuard stores transactions temporarily in a dictionary to correlate the components of each transaction. As transactions enter the mempool, we exclude irrelevant transactions to reduce time in identifying the target transactions. This reduces the overall identification time by 44.78% (Figure 7 compares the two methods). The rest of the transactions are then parsed and logged based on their function signatures and it creates a structured dictionary where each transaction hash is mapped to a series of potential exploit indicators. In a general sense, if a transaction contains a flash loan and token *transfers*, *approvals*, or *withdrawals*, then it gets flagged. This correlation is performed with a low detection time to ensure that suspect transactions are identified and marked for disruption before they are included in the next block on the blockchain.

**Disrupting Attack Atomicity.** Our method was able to disrupt the attack sequence and prevent the attacker from draining the funds from the protocol. We show counterattack disruption for the recent attacks in Table 7. We show detection time in Table 4 and time of attack disruption in Table 7. Finally, our results show that FlashGuard could have effectively rescued about $405.71 million in total losses if it were deployed prior to these attacks.

**Table 8:** Unseen evaluated vulnerabilities [51]

| Vulnerability | Description |
|---|---|
| Improper access control | Restricted function execution. |
| Unchecked return values | Tokens return false without reverting. |
| Uninitialized storage pointers | Corrupt state from uninitialized variables. |

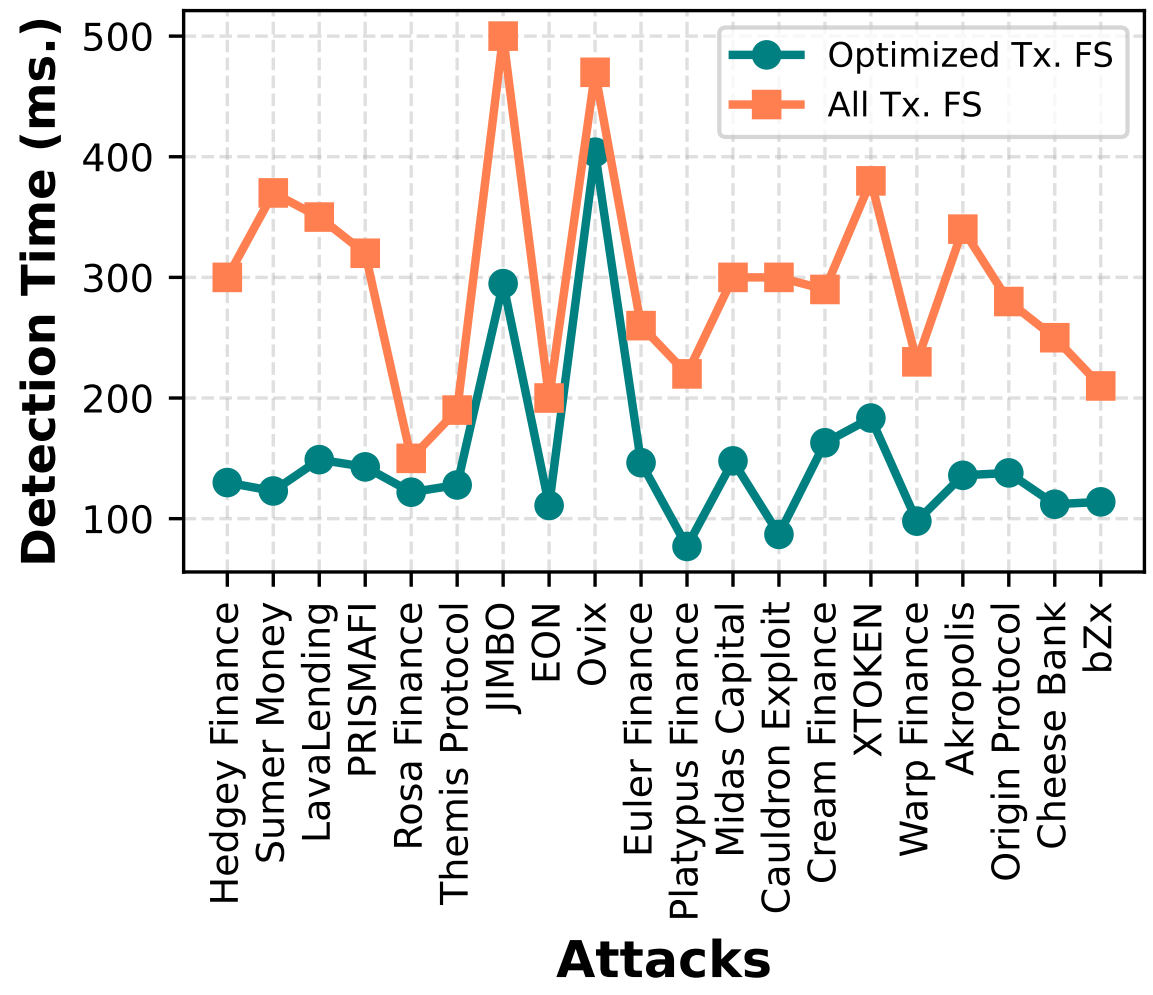

**Figure 7:** Detection times for non-price flash loan attacks using optimized function signatures (FS) and all function signatures, which includes all transactions in the mempool.

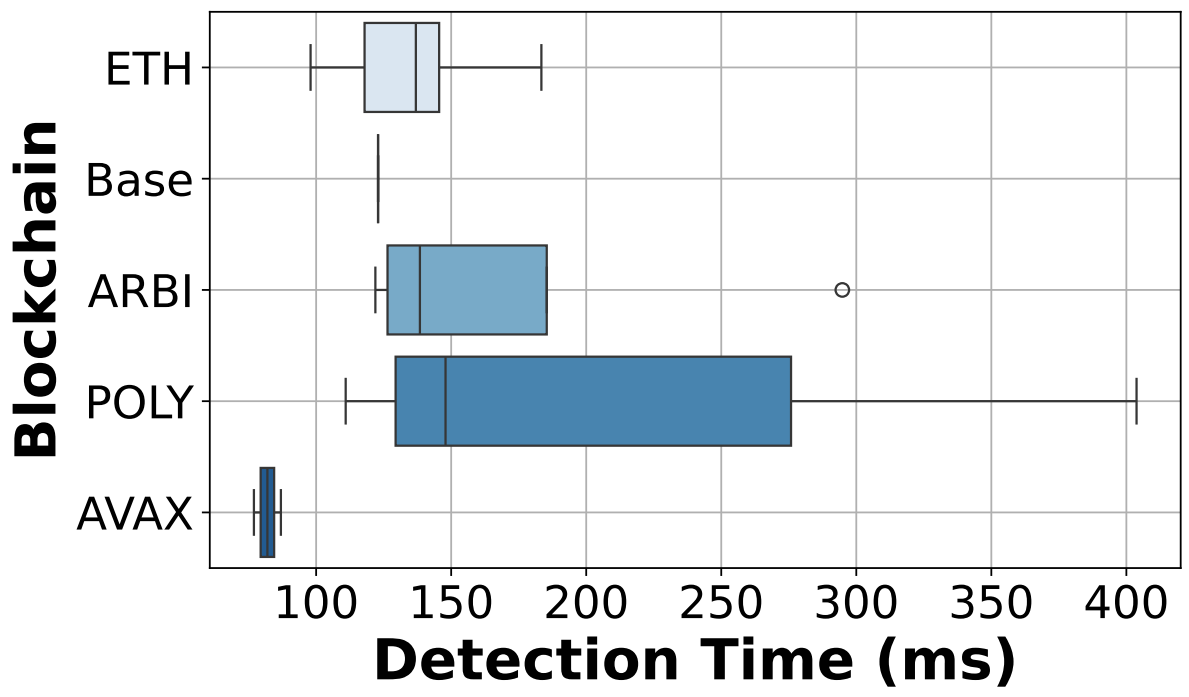

**Figure 8:** Detection time distribution across different blockchain networks.

Table 7 presents the attack disruptions using FlashGuard for various real-world attacks. FlashGuard achieves an efficient attack disruption with an average time of 410.92ms. This is considerably faster than other relevant studies, such as [59], which is considered state-of-the-art for synthesizing counterattacks. The median runtime overhead ranges from 0.08 to 0.59 seconds and the worst-case times range from 4.26 to 15.57 seconds. FlashGuard achieves this high performance by targeting the atomicity property of the blockchain to revert attack transactions while ensuring minimal overhead. We note that the average combined time for detection and disruption across the attacks is only 561.23 ms.

**Disrupting Unseen Attacks.** The evaluation of the FlashGuard against different unseen attack vectors is shown in Table 9. Specifically, we test six scenarios for access control, unchecked return value, and uninitialized storage attacks, as shown in Table 8. For platforms utilizing the ABSD model, FlashGuard successfully disrupts all attack types by maximizing the consumed gas to break the attack atomicity. This leads to the attack reversion as a result of FlashGuard's *dusting* counterattack that forces it to fail. On the other hand, platforms running under SFSD were protected against access control attacks, however, unchecked return value and uninitialized storage attacks proceed with no disruption. This is a result of FlashGuard minimizing the gas consumption for disruption in platforms with no pre-allowance granted. Nevertheless, FlashGuard still maintains protection against a critical vulnerability without incurring excessive resource costs. The gas cost trade-off shows that FlashGuard can achieve higher resource utilization for platforms under ABSD in exchange for better security and can reduce costs for platforms under SFSD with targeted protection measures.

**Table 9:** Disruption metrics for unseen attacks.

| Attack Type | Mechanism | Gas (Gwei) | Disrupted |
|---|---|---|---|
| Access Control Attack | ABSD | 1,714,375 | Yes |
| Access Control Attack | SFSD | 1,642,925 | Yes |
| Unchecked Return Value Attack | ABSD | 1,167,725 | Yes |
| Unchecked Return Value Attack | SFSD | 588,075 | No |
| Uninitialized Storage Attack | ABSD | 1,721,800 | Yes |
| Uninitialized Storage Attack | SFSD | 1,138,950 | No |

**Performance.** FlashGuard detects the attacks listed in Table 4 with an average latency of 150.31 ms, which allows passing the attack information discussed in Section 4.2 immediately to the disruption mechanism in real-time. On average, the disruption reverts attack transactions within 410.92 ms, with a worst-case time of 7.11 seconds. This is significantly faster than solutions like [59], which can take up to 15.57 seconds in the worst case. FlashGuard is also resource-efficient, using only 8.24% CPU and 0.40 MiB of memory for detection. This makes it scalable on EVM-compatible blockchains. With gas costs for countertransactions averaging 0.0038 ETH ($6.09), FlashGuard balances cost and effectiveness, aiming to provide practical and scalable protection against DeFi threats from non-price flash loan attacks.

# 6 RELATED WORK

Users utilize flash loans to borrow large amounts of cryptocurrency without collateral and flash loans require that the loan is repaid in the same block[53]. Even though it is used in many legitimate scenarios, such as arbitrage, maliciously attacks on flash loans can lead to huge financial losses. In non-price manipulation, flash loan is used to exploit vulnerabilities in smart contracts directly. Therefore, the main goal is to exploit the flaws in the protocol to drain funds rapidly, usually in a single transaction.

Several studies have investigated flash loan and its implications on DeFi. One of the previous works is FrontDef [22] in which the authors present a detection system designed to front-run malicious transactions, and monitor pending transactions in the mempool, and analyze the bytecode of the contracts involved. If a suspicious transaction is detected, FrontDef assembles and executes the same transactions to preempt the attack to prevent losses. This approach was validated against historical attacks and the detection and assembly of mimic transactions were successful in several cases. On the other hand, FlashSyn proposed by [14] synthesizes flash loan attacks using a counterexample-driven approximation. This tool evaluates DeFi protocols by simulating adversarial conditions to identify potential exploits. FlashSyn has demonstrated efficacy in detecting vulnerabilities across multiple DeFi platforms and provides insights into attack vectors and suggests mitigations. DeFiTail proposed in [34] focuses on cross-contract execution analysis. By examining smart contracts interactions, it identifies vulnerabilities and potential exploits that could be utilized by the adversarial, which includes flash loan attacks.

Most studies focus on analyzing flash loan attacks, particularly price manipulation attacks. Although existing solutions, such as [55, 56], demonstrate potential in mitigating price manipulation attacks, their practicality is limited to detecting such attacks. Generalized systems, such as [22, 59], attempt to front-run malicious transactions by increasing the gas cost to prioritize transaction inclusion, but they do not explicitly address non-price attack patterns. In particular, some studies, such as [59], focus only on mitigation without detection capabilities, which reduces their practical effectiveness in real-world scenarios. Also, their proposed techniques reconstruct the attack transaction to simulate its behavior and rescue smart contract funds. We note that this approach is time-consuming for atomic attack transactions, where numerous entities in smart contracts interact in a single rapid transaction. Nevertheless, in real-world blockchain transactions, especially in attack-related transactions, adversaries increase the gas fee significantly to avoid such countermeasures. This could lead to a successful attack identification, but the attack transaction can be confirmed before the counter-transaction is dispatched. Recent work in [8] introduced the preliminary idea for addressing this problem. Building on that work, this paper provides a comprehensive solution, including theoretical methods, algorithmic enhancements, and a detailed evaluation of FlashGuard on diverse DeFi protocols for different blockchains.

# 7 FUTURE WORK

We identify several important directions for future work on FlashGuard. In real-world scenarios, FlashGuard can integrate into the DeFi protocol without the need to modify its infrastructure, as it does not require redeploying the smart contract to include the FlashGuard mechanism. Future work could deploy FlashGuard on-chain for an extended period to evaluate its performance. Future studies also could explore how attacks can be obfuscated and design possible mitigation strategies by developing techniques to detect obfuscation. For example, machine learning models or pattern recognition algorithms could help recognize obfuscated signatures and identify obfuscated attack signatures. Finally, collaborating with DeFi platforms to deploy and test FlashGuard in real-world scenarios can provide more insights for real-world deployments. In addition, implementing a feedback loop where the system learns from any detection failures can significantly improve its performance over time, which can further increase the reliability and accuracy of FlashGuard.

# 8 CONCLUSION

We have presented FlashGuard, a novel method for run-time detection and prevention of non-price flash loan attacks in DeFi platforms that is generalizable to any Ethereum Virtual Machine compatible blockchain. FlashGuard leverages direct communication with miners for transaction submission and employs *dusting* counterattack transactions to disrupt attack sequences and prevent financial losses.FlashGuard detects non-price flash loans by identifying a combination of attack patterns, which we found to be common among all historical attacks, and introduces two attack disruption models for on-chain protection against non-price flash loan attacks. Our evaluation of real-world historical and unseen attacks shows that FlashGuard is highly effective, with an attack detection accuracy of over 99.93%, an average detection time of 150.31ms, and a disruption time of 410.92ms. FlashGuard is practical, has a low overhead, and is faster than the state-of-the-art.